\documentstyle[12pt]{article}
\title{ On the renormalization-scheme dependence in
quantum field theory. } 
\author{ }
\date{ }
\begin{document}
\maketitle
\vspace{-2.1cm}
\begin{center}
{\large Anton V. Konychev} \\
{\sl Institute for High Energy Physics, Protvino*}
\end{center}
\insert\footins{\ \ {\footnotesize * \ \ On leave from Moscow Institute 
of Physics and Technology,
Dolgoprudny \\
\ \ \hspace*{0.7cm} E-mail:\ \ ASP9401@mx.ihep.su }}

\vspace{0.5cm}
\begin{abstract}
Using quantum electrodynamics as an example, a dependence of physical predictions
of quantum field theory in a finite perturbation theory order
on the choice of renormalization scheme is studied. It is shown that
On-Mass-Shell renormalization scheme is distinguished in quantum electrodynamics not
only due to the agreement of the theory predictions with experimental results but also 
due to a set of specific theoretical properties. Thus performing renormalization
procedure in other theories, it seems reasonable to use some of
On-Mass-Shell renormalization scheme prescriptions.
\end{abstract}

\vspace{0.75cm}
{\bf  Introduction}
\vspace{0.25cm}

\noindent It is known that renormalized expressions for 
the divergent Feynman diagrams in quantum field theory are defined ambiguously.
The ambiguities arise due to a freedom in the choice of renormalization scheme (RS),
e.g. the choice of subtraction point in momentum space in $MOM$-scheme, the choice of 
particular value for scale parameter $\mu$ in $\overline{MS}$-scheme, etc. These, ultraviolet,
ambiguities are fixed in quantum electrodynamics (QED) by the use of On-Mass-Shell
renormalization scheme (see e.g. \cite{LL} $\S 110$ where the physical arguments
in favor of this scheme are put forward). Such a distinguished scheme is seemingly absent
in quantum chromodynamics (QCD) because of the absence of free quarks and gluons 
(see \cite{In} $\S9$, $\S16$).
Usually one fixes the ultraviolet ambiguities in QCD putting scale parameter
$\mu$ in $\overline{MS}$-scheme 
to be equal to the typical energy value for the process in question (see e.g. \cite{rpp}).
It is not well-defined
quantity and one can not expect the high accuracy predictions working in such a manner.
Moreover, an estimation of the uncertainty in the theory predictions
is, in fact, quite arbitrary.

The aim of this work is, using QED as an example, to explore the influence of ultraviolet ambiguities
on the theory predictions and show that On-Mass-Shell renormalization scheme is distinguished
in QED not only due to the phenomenological reasons but also due to a number of 
specific theoretical properties. It seems reasonable to use some of
On-Mass-Shell renormalization scheme prescriptions in other theories, e.g. in QCD.

\pagebreak[4]
{\bf  General statements and definitions}
\vspace{0.25cm}

\noindent Let us start with some general statements and definitions.
Coefficient functions appearing in the expansion of S-matrix in powers of interaction
contain, in general, products of free propagators. Free propagators are the singular
distributions and their products can appear to be ill-defined on the whole space of basic 
functions. The corresponding integrals in momentum space, the convolutions of Fourier
transforms of free propagators, appear to be ultravioletly divergent in this case.
These products, or corresponding convolutions, need an additional definition. 
This procedure is non-unique, the choice of renormalization
scheme is equivalent to the choice of particular definition for these  products
(or for corresponding convolutions in momentum space).

Exact, i.e. containing all radiative corrections, Green functions calculated in two different
renormalization schemes are related to each other by the Dyson transformation:
\begin{equation}
\Gamma (p\mid e,m;C)=z(C,C')\Gamma (p\mid e',m';C')
\end{equation}
Here $p=(p_1, p_2, \cdots )$ are momentum arguments of Green function.
$e,e'$ and $m,m'$ are the coupling constants and masses in renormalization schemes $C$ and $C'$.
Renormalization scheme is determined by a set of real parameters: $C=(c_1, c_2, \cdots )$.
These parameters appear as uncertain coefficients in Taylor expansions of renormalized 
self-energies and vertex functions and reflect the ambiguity of renormalization procedure 
(see \cite{BSh} $\S27.2$, $\S27.4$, $\S28.2$, \cite{BShs} $\S24.1$; see also formulas (5) and
(11) in the following section).  Relation (1) implies that Green function calculated in
scheme $C$ with coupling constant $e$ and mass $m$ is equal as a function of momenta,
up to the factor $z(C, C')$,
to the one calculated in scheme $C'$ with coupling constant $e'$ and mass
$m'$. In other words, a change of renormalization scheme (a change of
$C=(c_1, c_2, \cdots )$ in fact) can be compensated by the change of
$e$ and $m$. The deduction of relation (1) can be found in $\S34$ \cite{BSh}
\footnote{
The parameters $m$ and $m'$ are called masses arbitrarily. In general, they are not equal to
the physical mass $M$, the pole position of renormalized propagator. Like any other observable
quantity, $M$ is a function of the theory parameters and renormalization scheme:
$M=M(e,m;C)$.}.

An important remark should be made. Relation (1) is valid for exact Green functions only.
It is not valid in any finite order of perturbation theory. This leads to the dependence
of physical predictions in a finite order of perturbation theory on the choice
of renormalization scheme. One can often meet with
a statement that calculation of higher-order
corrections reduces RS dependence. This is, in general, a wrong statement.
Conversely, higher-order corrections introduce additional uncertainties into 
a final answer.
Let one have calculated Green functions and S-matrix elements up to the n-th order
of perturbation theory so that all singular expressions in the second, third, \ldots $n$-th orders
are already defined. Obtaining renormalized expressions for the 
Feynman diagrams of the n+1-th order, 
one has to use the prescriptions adopted in low orders to define the divergent subgraphs of
these diagrams. This statement is equivalent to the following one in counterterm approach: 
subdivergencies of the $n+1$-th order diagrams are removed by the counterterms of low orders.
However, the expressions for Feynman diagrams of the n+1-th order contain their own divergencies.
These divergencies arise when all integration momenta tend to infinity. To remove these divergencies
one needs the $n+1$-th order counterterms. These counterterms are, of course, of the same operator type
as the counterterms of low orders but have independent coefficients
\footnote{ In coordinate representation one needs to define  additionally the expression for 
$S_{n+1}(x_1, \cdots x_{n+1}) = i^{n+1}\,$T$(\,${\it L}$(x_1) \cdots ${\it L}$(x_{n+1}) \,)$
when all the arguments are equal to each other (see \cite{BSh} $\S29.2$).}.
Thus, calculation of each 
additional perturbation theory order increases the number of free
renormalization parameters and, hence, the uncertainty of the total result.   

An apparent contradiction with relation (1), i.e. with 
the actual absence of RS dependence for exact
Green functions, is due to the fact that the passage 
to the limit of exact Green functions
is performed {\it formally} when relation (1) is obtained. The higher-order corrections partially 
reduce the uncertainties induced by low orders but introduce new uncertainties which are
partially compensated by the further higher-order corrections and so on. The validity
of relation (1) for the formally obtained exact Green functions containing {\it all}
radiative corrections does not mean that RS dependence is being reduced when increasing but still finite
number of higher-order corrections is being taken into account \nolinebreak[4]
\footnote{To illustrate the situation let us consider function f determined by the formal series:
$$f(a_1,\cdots a_n,\cdots)=a_1+(a_2-{a_1 \over 2})+(a_3-{a_2 \over 2}-{a_1 \over 4})+\cdots+
(a_n-{a_{n-1} \over 2}-\cdots-{a_{n-k} \over 2^k}-\cdots-{a_1 \over 2^{n-1}})+\cdots$$
Truncation of this series at any finite order leads to the non-trivial dependence of
function f on its arguments. However, being considered as the sum of actually infinite
series, it does not depend on them.}.

Nevertheless, relation (1) is useful for theoretical investigations. Using it, one can
obtain renormalization group (RG) equations and demand that Green functions of the theory
satisfy these equations while the terms of low orders in expansions of these Green functions
in powers of interaction coincide with the results of perturbation theory.
One can expect that such a RG-invariant Green functions will exhibit improved 
approximating properties. However, this procedure does not remove RS dependence since
using perturbative approximations calculated in different renormalization schemes to
construct RG-invariant expressions one obtains, in general, different RG-invariant
results (see \cite{BSh} $\S48.5$)
\footnote{
Recall that the coefficient of term proportional $\alpha_s^4$ in the $\beta$-function
expansion in powers of $\alpha_s$ in massless QCD depends on the choice of
renormalization schemes. If quark masses are not equal to zero, even the coefficient of term
proportional $\alpha_s^2$ is RS-dependent. As a consequence, the expression for invariant
coupling constant also appears to be RS-dependent.
}.
Thus  RG invariance is not equivalent to  RS independence. The only way to avoid
uncertainties in such a situation is to find convincing arguments in favor of the particular
renormalization prescription.

S-matrix elements are obtained from Green functions by means of reduction formula:
\begin{eqnarray}
\nonumber & S(p_1 \cdots p_n; p_1' \cdots p_m')=\prod_{1}^{n}{ \lbrace {z_k}^{-1/2} f^+ (p_k)
\Delta^{-1} (p_k) \rbrace} \prod_{1}^{m}{ \lbrace {z_l}^{-1/2} f^- (p_l') \Delta^{-1} (p_l')
\rbrace} \times & \\ & \times G^{(m+n)}(p_1 \cdots p_n; p_1' \cdots p_m' ) &
\end{eqnarray}
(see e.g. \cite{FSl} ch.4 $\S9$).
Here $f^-$ and $f^+$ are wave functions of initial and final particles, $\Delta^{-1} (p)$
is the inverse free propagator with physical mass, e.g. $\Delta^{-1} (p)=p^2-M^2$ for
the scalar particle. Factor $z^{-1/2}$ is the square root of the pole residue of
renormalized propagator, $z=(p^2-M^2)G^{(2)}(p)\mid_{p^2=M^2}$ for the scalar particle.
The presence of  $z^{-1/2}_{k,l}$ factors is an important 
peculiarity of formula (2).
Their origin is related to the fact that when $z\ne 1$, wave functions of external
particles are effectively renormalized. A probability of transition from the initial
state to the final one is related to the S-matrix element by the formula
\begin{equation}
W_{i\rightarrow f}={ {\vert \langle \Phi_f \mid S \mid \Phi_i \rangle \vert}^2 \over
\langle \Phi_f \mid \Phi_f \rangle \langle \Phi_i \mid \Phi_i \rangle }
\end{equation}
It is convenient to remove factors $z^{1/2}_{k,l}$ from the denominator of formula (3) to reduction
formula (2). For wave functions $f^-$ and  $f^+$ in formula (2) 
conventional normalization remains.

\vspace{0.5cm}
{\bf RS dependence in perturbation theory}
\vspace{0.25cm}

\noindent Let us turn now to a detailed consideration of the RS dependence in finite order
of perturbation theory. As an example, the scattering of an electron from the external source 
in the third order of expansion in powers of interaction constant and in the first one
of external field will be
considered (see also \cite{Shv} ch.15 $\S3$). Being simple enough, 
this example makes it possible 
to observe in detail the influence of ultraviolet ambiguities on the physical predictions of
the theory. Diagrams corresponding to the process are shown in the figure 1.

\vspace{0.5cm}
{\sf Renormalizition of electron mass}
\vspace{0.25cm}

\noindent Diagrams c) and d) make the contribution to the renormalization of the 
elektron mass parameter. The following expression corresponds to diagram c)
\footnote{ Factor $2\pi i$ common for diagrams a) --- e) is omitted.}:
\begin{equation}
M_c=\stackrel{-}{u}(p_f)e\gamma^{\mu}A_{\mu}{1 \over \hat p_i - m}\Sigma(p_i)u(p_i)
\end{equation}
where
$$\Sigma(p)={\alpha \over 4\pi^3 i} \int_{ }^{ }{ \gamma^{\nu} {\hat p - \hat k +m \over
{(p-k)}^2 - m^2 + i\epsilon} \gamma_{\nu} {dk \over k^2 + i\epsilon} }$$
Analogous expression corresponds to diagram d). $\Sigma(p)$, the electron self-energy,
is determined up to two real coefficients (see \cite{BSh} $\S27.2$, $\S35.2$):
\begin{equation}
\Sigma(p)=c_0+c_1(\hat p - m)+\tilde \Sigma(p)
\end{equation}
One can choose the following normalization for $\tilde \Sigma(p)$:
\begin{equation}
\tilde \Sigma(m)=0, \qquad {d \tilde \Sigma(p) \over d\hat p} \bigg \vert_{\hat p = m}=0
\end{equation}
where ${d \tilde \Sigma(p) \over d\hat p} = {\partial \tilde \Sigma(\hat p, p^2) \over \partial \hat p}
+ 2 \hat p {\partial \tilde \Sigma(\hat p, p^2) \over \partial p^2}$ \ 
\footnote{$\Sigma(p)$ contains an infrared regulator. The general form for $\Sigma(p)$ is
$\Sigma(\hat p, p^2) = F_0(p^2)+\hat p F_1(p^2)$ where $F_0$ and $F_1$ are the some functions.
Expression (5) is equivalent to the fact 
that $F_0$ and $F_1$ are known up to the independent of $p^2$ 
additive values. Renormalization scheme is determined by the particular choice of $c_0$ and
$c_1$. It should be recognized that in perturbation theory order being considered
the expression for finite part of electron self-energy
obtained in {\it any} renormalization scheme ($\overline{íS}$, $MOM$, etc.) 
can be reduced to
the form (5) where $\tilde \Sigma(p)$ is normalized according to, e.g. (6). }.
Let us substitute (5) into expression (4):
\begin{displaymath}
 M_c=\stackrel{-}{u}(p_f)e\gamma^{\mu}A_{\mu}{c_0 \over \hat p_i - m}u(p_i)+
\stackrel{-}{u}(p_f)e\gamma^{\mu}A_{\mu}{c_1 \over \hat p_i - m}(\hat p_i - m)u(p_i)+
\end{displaymath}
\begin{equation}
+\stackrel{-}{u}(p_f)e\gamma^{\mu}A_{\mu}{ 1 \over \hat p_i - m}\tilde \Sigma(p_i)u(p_i)
\end{equation}
If one supposes that $u(p_i)$ satisfies the equation $(\hat p_i - m)u(p_i)=0$,
the first term in the r.h.s. of equation (7) will contain a singularity when $c_0 \ne 0$.
One can not calculate the total transition amplitude by the direct summation of diagrams
a) --- e) if $c_0 \ne 0$.
However, the $c_0 \ne 0$ condition can be treated in the following manner.
Let us consider electron propagator containing appropriate radiative corrections:
\begin{displaymath}
G^{(2)}={ 1 \over \hat p - m}+{ 1 \over \hat p - m} \Sigma(p)
{ 1 \over \hat p - m} = { 1 \over \hat p - m} \Biggl ( 1+ { \Sigma(p) \over \hat p - m} \Biggr )
\approx { 1 \over \hat p - m}\, \Biggl ( { 1 \over  1- {\Sigma(p) \over \hat p - m}  } \Biggr ) = 
\end{displaymath}
\begin{equation}
={ 1 \over \hat p - m -\Sigma(p)}
= { 1 \over \hat p - m - c_0 -c_1(\hat p - m)-\tilde \Sigma(p)}
\end{equation}
One sees from the r.h.s. of equation (8) that if $c_0 \ne 0$, the denominator is not equal to zero
when $\hat p=m$. Thus, parameter $m$ no longer coincides with physical mass $M$.
The $u(p)$ should satisfy the equation $(\hat p - M)u(p)=0$.
To obtain the total transition amplitude when $c_0 \ne 0$ one should calculate vertex function 
$\Gamma^\mu = \gamma^\mu + \Lambda^\mu (p_i,p_f;m)$,
propagators $G^{(2)}$ (see equation (8)) and $D^{(2)}_{\mu \nu}$
\footnote{ Function $\Pi(q^2;m)$ is related to the polarization operator $\Pi^{\mu \nu}(q^2;m)$
by the formula $\Pi^{\mu \nu}(q^2;m)=(g^{\mu \nu}-{q^\mu q^\nu \over q^2})\,q^2\,\Pi(q^2;m)$.
The explicit expressions for $\Pi(q^2;m)$ and $\Lambda^{\mu}(p_i,p_f;m)$ see in 
$\S27,28$ \cite{BSh}. }:
\begin{equation}
D^{(2)}_{\mu \nu}=\biggl ( g_{\mu \nu}-{q_\mu q_\nu \over q^2} \biggr )\,{1 \over q^2- q^2\Pi(q^2;m)}
+{1 \over q^2} {q_\mu q_\nu \over q^2}
\end{equation}
Then one should
construct full (connected) Green function $ G^{(3)}=G^{(2)}\,e\,\Gamma^\mu G^{(2)} D^{(2)}_{\mu\nu}$
and use reduction formula (2). Thus one obtains
\begin{equation}
M_{total}=\stackrel{-}{u}(p_f)e\, \Bigl (\gamma^\mu+ \Lambda^\mu (p_i,p_f;m) \Bigr )
u(p_i) {1 \over 1- \Pi(q^2;m) }A_\mu
\end{equation}
Factors arising when the pole residues of propagators are different from unity are temporarily
omitted in formula (10). It is important here that $\Lambda^{\mu}(p_i,p_f;m)$ and $\Pi(q^2;m)$
depend on parameter $m$ rather than physical mass $M$.
Thus, the total result contains the dependence on arbitrary parameter $m$.
In contrast to the case of exact
transition amplitude, 
a change of parameter $m$ in expression (10) can not be compensated 
by the change of renormalization scheme. A predictive power of the theory is lost.  

If one puts $c_0=0$ (the requirement of On-Mass-Shell renormalization scheme), 
calculations can be performed both by
the first method (the direct summation of diagrams a) --- e) ) and by the second one (the 
construction of full Green function followed by the application of reduction formula (2) ).
The results agree with each other up to the higher-order corrections. 

It is to be noted
that imaginary parts of $\Pi(q^2;m)$ and $\Sigma(p;m)$ are ultravioletly finite. Imaginary part
of $\Pi(q^2;m)$ is different from zero when $q^2 > 4m^2$. Thus the threshold for real
electron-positron pair production is determined by the parameter $m$ which, hence, should
be equal to physical mass $M$. In this case 
the renormalized S-matrix will be unitary in each
perturbation theory order. Otherwise, the exact S-matrix only will be unitary whereas in
finite orders the unitarity will be broken.

\vspace{0.5cm}
{\sf Renormalization of electron propagator pole residue}
\vspace{0.25cm}

\noindent Let us turn now to the consideration of the second term in the r.h.s. of equation (7).
In what follows, to avoid difficulties arising when $m \ne M$,  $c_0$ in expression (5)
will be put to be equal
to zero, i.e. $\Sigma(m)=0$. The second term in r.h.s. of equation (7) contains an
uncertainty: one can act by the operator $(\hat p_i - m)$ on the spinor $u(p_i)$ and obtain
zero, on the other hand one can "cancel" operators $(\hat p_i - m)$ in the numerator and the
denominator and obtain the expression $c_1\stackrel{-}{u}(p_f)e\gamma^{\mu}A_{\mu}u(p_i)$.
To resolve this uncertainty one can use the adiabatic hypothesis (for more details, see
\cite{Shv} ch.15 $\S3$). As a result one obtains for the second term the expression
${1 \over 2}c_1\stackrel{-}{u}(p_f)e\gamma^{\mu}A_{\mu}u(p_i)$
\footnote{ The question of the external lines renormalization of Feynman diagrams is not
simple. See in this connection \cite{MP}. See also the end of $\S38$ \cite{BD}.}.
Factor ${1 \over 2}$ is of great importance. Owing to gauge invariance of
quantum electrodynamics,
$\Lambda^\mu$ and $\Sigma$ are related by the Ward identity:
$$
\Lambda^\mu (p,p) = - {\partial \Sigma(p) \over \partial p_\mu}
$$
An ambiguous renormalization coefficient $c_2$ in $\Lambda^\mu$ is therefore 
not independent of $c_1$. If one writes $\Lambda^\mu$ as
\begin{equation}
\Lambda^\mu (p_i,p_f)=\tilde \Lambda^\mu (p_i,p_f) - c_2\gamma^\mu
\end{equation}
where
\begin{equation}
\stackrel{-}{u}(p) \tilde \Lambda^\mu (p,p) u(p) = 0
\end{equation}
then $\tilde \Lambda^\mu (p,p) = - {\partial \tilde \Sigma(p) \over \partial p_\mu}$  where
$\tilde \Sigma(p)$ is normalized according to (6) and, hence, $c_2=c_1=c$. Taking into
account factor ${1 \over 2}$ for diagrams c) and d) and calculating the sum of
b)\footnote{ Expression $\stackrel{-}{u}(p_f)\,e\,\Lambda^\mu (p_i,p_f)u(p_i)A_\mu$ 
corresponds to diagram b). }, c) and d), one sees that terms containing $c$ cancel
from the total amplitude. Thus the total result is free from the ambiguities
connected with renormalization
of the vertex function and the electron propagator pole residue
\footnote{ In QCD coefficient $c_1$ also cancels from the total result though a simple
relation $c_2=c_1$ is not valid there (see \cite{In} $\S 9.2$).
The choice of particular normalization for $c_1$ is immaterial when one works in the first method.
}.

If one uses the second method to obtain the total amplitude, the following 
expression is arrived at:
\begin{equation}
M_{total}
=\stackrel{-}{u}(p_f)\sqrt{{1 \over 1-c}}\,e\, \Bigl (\gamma^\mu+\tilde \Lambda^\mu (p_i,p_f)-
c\gamma^\mu \Bigr ) \sqrt{{1 \over 1-c}}\,u(p_i) {1 \over 1- \Pi(q^2) }A_\mu
\end{equation}
Let us rewrite (13) as
\begin{equation}
M_{total}=\stackrel{-}{u}(p_f)\,e\, \Bigl (
\gamma^\mu +{1 \over 1-c} \tilde \Lambda^\mu (p_i,p_f)\, \Bigr )\,u(p_i)
{1 \over 1- \Pi(q^2) }A_\mu
\end{equation}
Due to relation (12), $M_{total}$ is independent of $c$ when $q^2=0$ ($p_i=p_f=p$). However,
when $q^2 \ne 0$, $M_{total}$ depends on $c$. It is quite a 
disagreeable fact. Coefficient
$c$ can depend on an infrared regulator and gauge parameter, 
e.g. this is the case if one
normalizes $\Sigma(p)$ as follows: $\Sigma(m)=0$, 
${\partial \Sigma(p) \over \partial \hat p} \big \vert_{\hat p = m}\!\!=0$ 
(see \cite{BSh} $\S35.2$; one should take into account that normalization for $\tilde \Sigma(p)$
and, hence, the definition of coefficient $c_1$ in this paper are different from those in the
reference).
In $\overline{MS}$-scheme $c$ depends on an infrared regulator, gauge parameter and
scale parameter $\mu$. Thus, in general, expression (14) leads to an 
incorrect result.
However, if one puts $c=0$ in (14) (On-Mass-Shell renormalization scheme),
the result will coincide, up to the higher-order corrections, 
with the one obtained
in the first method where there is no ambiguity due to coefficient $c$
\footnote{ In the pure transversal gauge (Landau gauge) ultraviolet divergencies related to
the coefficients $c_1$ and $c_2$ are absent. If one writes the expression 
similar to (5)
for $\Sigma(p)$, the coefficient before $(\hat p - m)$ will not be uncertain. It will depend on
an infrared regulator. When one uses the first method, this coefficient cancels from the
sum of diagrams b), c) and d) and, hence, from the total amplitude. (Of course,
vertex
function contains some other terms depending on infrared regulator.) Working in the second 
method, one has to demand that this coefficient should
be equal to zero to obtain a
reasonable result despite the absence of a freedom in its normalization in Landau gauge.
A detailed consideration of infrared divergencies is beyond the scope of this paper.
It should be still emphasized once again 
that if one works in the first method, one has no need
to take care of the particular normalization for coefficient $c_1$.}.

\vspace{0.5cm}
{\sf Renormalization of photon propagator pole residue}
\vspace{0.25cm}

\noindent Finally, let us turn to the consideration of the ambiguity connected with
normalization of the photon propagator pole residue. One can write $\Pi(q^2)$ as
$\Pi(q^2)=c_3+\tilde \Pi(q^2)$ where $\tilde \Pi(0)=0$
\footnote{ As in the case of $\Sigma(p)$, this is a 
general expression for $\Pi(q^2)$
explicitly distinguishing the  ambiguity contained in it.}.
In exact Green functions and S-matrix elements
a change of $c_3$ can be compensated by the change of $e$. This is
not the case in the finite order of perturbation theory. Let us consider the expression
for the total amplitude obtained in the second method. To take into account the charge
renormalization correctly, 
one should write $A_\mu$ as $A_\mu=D_{\mu \nu}\,e\,J_\nu^{ext}$
where $J_\nu^{ext}$ is the external current responsible for field $A_\mu$,
$D_{\mu \nu}$ is free photon propagator. The expression for the total amplitude is
\begin{equation}
M_{total}=\stackrel{-}{u}(p_f)\,e\, \Bigl (\gamma^\mu+ e^2 \bar \Lambda^\mu (p_i,p_f)\, \Bigr )\, u(p_i)
{1 \over 1- c_3-e^2 \bar \Pi(q^2) }\,e\,J_\mu^{ext}
\end{equation}
Factors $e^2$ contained in $\tilde \Lambda^\mu (p_i,p_f)$ and $\tilde \Pi(q^2)$ are
explicitly distinguished in formula (15). One can rewrite (15) as
\begin{equation}
M_{total}=\stackrel{-}{u}(p_f) \Bigl (\gamma^\mu+ e^2 \bar \Lambda^\mu (p_i,p_f)\, \Bigr )\, u(p_i)
{1 \over {1- c_3 \over e^2} - \bar \Pi(q^2) } J_\mu^{ext}
\end{equation}
If one were allowed to neglect term $e^2 \bar \Lambda^\mu (p_i,p_f)$ in (16),
the expression for the total amplitude would possess the desired property. A change of $c_3$
would be compensated by the change of $e$. However, it is this term that 
is the dominant
radiative correction when $-q^2 \gg m^2$ 
(see e.g. \cite{LL} $\S122$) so that it can not be 
ignored.

Working in the first method and choosing the appropriate value for coefficient $c_3$,
one can make a contribution to the total amplitude from diagram e)
at some $q^2 \ne 0$ to be equal to zero. However, there is little sense in doing so because,
as it has been just noted, the sum of diagrams b), c) and
d) yields the major contribution to the total amplitude when $-q^2 \gg m^2$
rather than diagram e).

The value $c_3=0$ (i.e. $\Pi(0)=0$, On-Mass-Shell renormalization scheme) is distinguished. In this case
{\it all} the radiative corrections 
to scattering process in question tend to zero when $q^2 \rightarrow 0$
and the process is described with tree diagram a) only.
Recall that the Compton scattering of the
photon by the electron in the limit of low-energy photon is also described with the
tree diagrams only in this scheme. Thus the numerical value of electric charge can be easily 
extracted from the experiment
\footnote{ It is to be noted that if the
numerical value of a coupling constant is extracted from large-$q^2$ experimental
data where higher-order corrections are significant, one needs to verify the stability
of this value when the number of 
perturbation theory orders taken into account is changed.
Such a situation is typical in QCD where low-$q^2$ domain is experimentally inaccessible.
The last fact does not mean, of course, 
that On-Mass-Shell renormalization scheme can not 
be applied there.}.

S-matrix calculated in On-Mass-Shell renormalization scheme is
unitary in each perturbation theory order whereas in other schemes the exact S-matrix only
is unitary. An important property of the scheme is the absence of radiative
corrections to external lines of Feynman diagrams.
Effects of self-interaction already taken into account by the initial approximation no longer
occur in the theory in this scheme. The first and the second methods of obtaining the
transition amplitude are equivalent in finite orders of perturbation theory in 
On-Mass-Shell renormalization scheme only.

\vspace{0.5cm}
{\sf Additional remarks}
\vspace{0.25cm}

\noindent It has been demonstrated above 
that due to gauge structure of quantum electrodynamics
an uncertain parameter arising when renormalization of 
the vertex function is performed is
related to the appropriate one of electron self-energy.
If one has fixed renormalization prescription for the electron self-energy,
the prescription for the vertex function has been also fixed.
In a theory without gauge symmetry, e.g. $\lambda \phi^4$-theory, uncertain
renormalization parameter of 4-point  vertex function is independent of renormalization
parameters of self-energy. When one has defined the
renormalized self-energy according to the requirements of On-Mass-Shell renormalization scheme,
a freedom in normalization for the vertex function can be used to make
radiative corrections at the point where the interaction constant is measured to be equal to zero. 
In fact, it is possible in $\lambda \phi^4$-theory at the point $s=4m^2$, $t=u=0$ (threshold point)
only. At other points in physical domain an 
imaginary part of the vertex function is different 
from zero (see \cite{BShs} appendix 7, $\S24.1$). Thus, there is 
a distinguished renormalization
scheme in $\lambda \phi^4$-theory too
\footnote{ If one uses another renormalization prescription for 4-point vertex function
in $\lambda \phi^4$-theory, a numerical value of the interaction constant corresponding 
to this prescription can
be related to the threshold value of the interaction constant (see \cite{BSh} $\S36.2$). However, the
vertex functions obtained in two different renormalization schemes {\it functionally}
differ from one another so that their numerical values at points distinct from the point
$s=4m^2$, $t=u=0$ (and, hence, physical predictions of the theory) are different.}.   

\vspace{0.5cm}
{\bf Conclusion}
\vspace{0.25cm}

\noindent Let us summarize the main statements of this paper.

Renormalized expressions for the divergent Feynman diagrams in quantum field theory are defined
ambiguously. Physical predictions in a finite order of perturbation theory depend on the
choice of renormalization scheme. Calculation of higher-order corrections does not lead, in general,
to the diminishing of  renormalization-scheme dependence.

On-Mass-Shell renormalization scheme is distinguished in quantum electrodynamics
among other possible ones not only
due to the phenomenological reasons but also due to a number of specific theoretical properties.
Thus it seems reasonable to use some of On-Mass-Shell renormalization scheme prescriptions
performing renormalization procedure in other theories.

\vspace{0.5cm}
{\bf Acknowledgements}
\vspace{0.25cm}

\noindent The author is grateful to Yu.F.Pirogov and V.V.Kabachenko for useful discussions and
criticism.  Program \cite{RS} has been used to draw Feynman diagrams. This work is supported
in part by the Russian Foundation for Basic Research (project 96-02-18122a) and in part by the
Competition Center for Fundamental Natural Sciences (project 95-0-6.4-21).

\vspace{0.5cm}

%
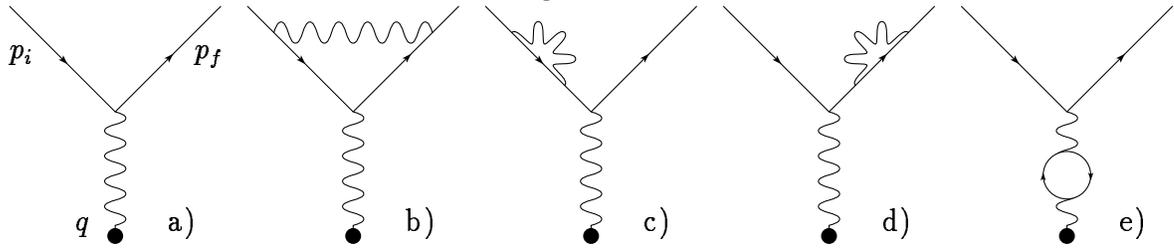
\end{document}